\documentclass[aps,pra,reprint,superscriptaddress,secnumarabic,amssymb]{revtex4-1}

\newcommand{\E}{\text{E}}

\newcommand{\Sa}{\hat{S}_s}
\newcommand{\Sad}{\hat{S}_s^\dagger}

\newcommand{\Wa}{\hat{W}_w}
\newcommand{\Wad}{\hat{W}_w^\dagger}
\newcommand{\U}{U_\tau}
\newcommand{\Ud}{U_\tau^\dagger}

\newcommand{\x}{\hat{x}}
\newcommand{\p}{\hat{p}}
\newcommand{\Ha}{\hat{H}}

\newcommand{\fref}[1]{Fig.\hspace{0.025in}\ref{#1}}
\newcommand{\sref}[1]{Sec. \ref{#1}}
\newcommand{\eref}[1]{Eq.\hspace{0.025in}(\ref{#1})}

\newcommand{\Real}{\operatorname{Re}}

\newcommand{\pref}[1]{(\ref{#1})}

\newcommand{\abs}[1]{\left| #1 \right|} 
\newcommand{\ket}[1]{\left| #1 \right\rangle} 
\newcommand{\bra}[1]{\left\langle #1 \right|} 
\newcommand{\braket}[2]{\left\langle #1 \vphantom{#2} \right|
\left. #2 \vphantom{#1} \right\rangle} 

\usepackage{graphicx}
\usepackage[normalem]{ulem}
\usepackage{color}
\usepackage{amsmath}
\usepackage{enumitem}

\begin{document}
\title{Robust Weak-Measurement Protocol for Bohmian Velocities}

\author{F. L. Traversa}
\email{fabiolorenzo.traversa@uab.cat}
\affiliation{Departament d'Enginyeria Electr\`onica, Universitat Aut\`onoma de Barcelona, 08193-Bellaterra (Barcelona), Spain}
\affiliation{Department of Physics, University of California, San Diego, La Jolla, California 92093, USA}
\author{G. Albareda}
\affiliation{Departament d'Enginyeria Electr\`onica, Universitat Aut\`onoma de Barcelona, 08193-Bellaterra (Barcelona), Spain}
\author{M. Di Ventra}
\affiliation{Department of Physics, University of California, San Diego, La Jolla, California 92093, USA}
\author{X. Oriols}
\email{xavier.oriols@uab.cat}
\affiliation{Departament d'Enginyeria Electr\`onica, Universitat Aut\`onoma de Barcelona, 08193-Bellaterra (Barcelona), Spain}

\begin{abstract}
We present a protocol for measuring Bohmian - or the mathematically equivalent hydrodynamic - velocities based on an ensemble of two position measurements, defined from a Positive Operator Valued Measure, separated by a finite time interval. The protocol is very accurate and robust as long as the first measurement uncertainty divided by the finite time interval between measurements is much larger than the Bohmian velocity, and the system evolves under flat potential between measurements. The difference  between the Bohmian velocity of the unperturbed state and the measured one is predicted to be much smaller than $1 \%$ in a large range of parameters. Counter-intuitively, the measured velocity is that at the final time and not a time-averaged value between measurements.
\end{abstract}

\maketitle

\section{Introduction}

The velocity of a classical object, requiring two position measurements, is trivially implemented in many apparati which control our daily activity.  On the contrary, in the quantum world, such measurements are much more complicated. The first position measurement implies a perturbation on the quantum system so that the knowledge of the velocity without perturbation is hardly accessible. One can minimize the back-action of the measurement on the system using weak measurements. Such measurements were initially developed by Aharonov, Albert and Vaidman (AAV) \cite{AAV} more than two decades ago and they are receiving increasing attention \cite{weak5,Wiseman, weak2,weak4,Bohmphoton,weak3,Hiley,Vaidman,Garretson} nowadays. As a relevant example, the spatial distribution of \emph{velocities} of relativistic photons in a double slit scenario has been measured, and the associated quantum trajectories reconstructed \cite{Bohmphoton}. However, we may ask the question: \emph{Does the ensemble velocity obtained from weak measurements have a clear physical meaning?} A partial answer was provided recently by Wiseman \cite{Wiseman}. Using the weak AAV value \cite{AAV}, he showed that the ensemble velocity constructed from an arbitrarily pre-selected state and a post-selected position eigenstate, with an infinitesimal temporal separation between position measurements, exactly corresponds to the Bohmian velocity \cite{Bohm} of the unperturbed state. Note that Wiseman's answer is only valid for non-relativistic scenarios (thus, strictly speaking, excluding \cite{Bohmphoton}).

We emphasize that two weak position measurements on an individual state do not provide the Bohmian velocity because of the unavoidable back-action \cite{Golstein}. However, for an idealized scenario,  Wiseman showed that when the individual measurements are repeated over an ensemble of identical states, the final ensemble velocity is identical to the Bohmian velocity of the unperturbed state  \cite{Wiseman}. These ensemble velocities can be interpreted either as the orthodox hydrodynamic velocity \cite{Madelung,Di_Ventra} or as a genuine measurement of the Bohmian velocity \cite{Golstein}. Following the recent literature \cite{Bohmphoton,Wiseman,Golstein}, we will refer to these ensemble velocities as Bohmian velocities, however the adjectives \emph{Bohmian} and \emph{hydrodynamic} are fully interchangeable in this work.

The practical conditions for measuring Bohmian velocities in a laboratory are different from the idealized theoretical scenario studied by Wiseman \cite{Wiseman} (implying discrepancies between the measured velocity and the expected one). First, {\it weak} measurements in a laboratory can be outside the linear-response regime assumed in the AAV development \cite{nori}. Second, position measurements have a small but finite uncertainty, meaning that the post-selected state is not an exact position eigenstate. Third, the time-separation between measurements must be finite. In this paper we bring the original Wiseman's conclusions about the measurement of Bohmian velocities into practical laboratory conditions, free from previous idealized assumptions. We will use the Positive Operator Valued Measure (POVM) framework \cite{nori} (instead of the AAV value) allowing positions uncertainties in both measurements and we will consider a finite time interval between position measurements.

\section{Ensemble velocity}

\subsection {Definition of ensemble velocity}

>From a large set of measured positions, $x_w$ at time $t_w$ and $x_s$ at $t_s=t_w+\tau$,  we construct the experimental velocity as:
\begin{equation}
\label{vexp}
v_{e}(x_s,t_s)=\frac { E[(x_s - x_w)|x_s]} {\tau},
\end{equation}
being $E[(x_s - x_w)|x_s]$ the ensemble average of the distance  $x_s-x_w$, conditioned to the fact that $x_s$ is effectively measured.
Since $E[x_s|x_s]=x_s$, the theoretical computation of the velocity $v_e$ does only require evaluating  $\E[x_w | x_s]$ using standard probability calculus,
\begin{equation}
\label{cond}
\E[x_w | x_s] = \frac {\int dx_w  x_w P(x_w\cap x_s)} {P( x_s)},
\end{equation}
with $P(x_w \cap x_s)$ the joint probability of the sequential measurements of $x_w$ and $x_s$, and $P(x_s)$ of $x_s$. After properly modeling the system perturbation due to the measurement, both probabilities can be computed.

\subsection{Two consecutive  POVMs separated by a finite time interval}

The POVM appears as a natural modeling of a measuring process \cite{weak} when the laboratory is divided into the quantum system and the rest (including the measuring apparatus).
Thus, the perturbation of the state due to the measurement of the first position $x_w$ can be defined  through POVMs. In this treatment we chose the Gaussian measurement Krauss operators
\begin{eqnarray}
\label{W}
\Wa=C_w  \int dx e^{-\frac {(x_w-x)^2} {2\sigma_w^2}} \ket x \bra x,
\end{eqnarray}
where $\sigma_w$ is the experimental uncertainty. The measured position $x_w$ belongs to the set $\mathfrak{M}$ of all possible measurement outputs of the apparatus. For simplicity, we assume $\mathfrak{M}\equiv\mathbb{R}$ in a 1D system, being the extension to the 3D spatial domain straightforward. Then, the normalization coefficient $C_w=(\sqrt\pi\sigma_w)^{-1/2}$ is fixed by the condition $\int dx_w \Wad \Wa=I$. Due to the unavoidable uncertainty on any position measurement, we consider an equivalent operator for the second position measurement of $x_s$:
\begin{eqnarray}
\label{S}
\Sa=C_s  \int dx e^{-\frac {(x_s-x)^2} {2\sigma_s^2}} \ket x \bra x.
\end{eqnarray}
We remark here that the choice of Gaussian measurement operators is not the only possible one that leads to our results. In fact, it can be proven that any POVM that weakly perturbs the wave function only in a neighborhood of $x_w$ ($x_s$) of radius $\sigma_w$ ($\sigma_s$), and \emph{cancels} the wave function in any other position leads to equivalent results. Thus the choice of Gaussian POVM is purely formal. It allows a simple analytical treatment.
Now, using the definitions in \pref{W} and \pref{S}, we can compute $P(x_w \cap x_s)$ and $P(x_s)$ from the Born rule, as:
\begin{align}
\label{prob1}
P(x_w \cap x_s)=&\bra{\Psi}\Wad\Ud\Sad\Sa\U\Wa\ket{\Psi} \\
\label{prob2}
P(x_s)=&\int dx_w P(x_w \cap x_s).
\end{align}
being $\ket {\Psi(t_w)}\equiv \ket {\Psi}$ the initial state. Strictly speaking, contrarily to the AAV expression \cite{AAV}, we are using a weak measurement without post-selection. The final state of the system (determined by the time-evolution of the initial state $\ket {\Psi}$ and the measurement processes) has no relevant effect when computing \pref{prob1} and \pref{prob2}.


\subsection{Calculation of the ensemble velocity}
\label{sec_velocity}

Let us now analyze $P(x_s)$ in detail by substituting \eref{W} and \pref{S} into \eref{prob2}. Then, we have
\begin{multline}
\label{Pxs}
P(x_s)=C_w^2 \iiint  dx_wdx\rq{}dx\rq{}\rq  e^{-\frac{(
x_w-x\rq)^2}{2\sigma_w^2}} e^{-\frac{(x_w-x\rq{}\rq)^2}{2\sigma_w^2}}\times\\
\times \braket {\Psi} {x\rq} \bra{x\rq}\Ud\Sad\Sa\U\ket{x\rq{}\rq} \braket {x\rq{}\rq} {\Psi}.
\end{multline}
Integrating over $x_w$ and using \eref{S}, we can rewrite \eref{Pxs} as:
\begin{multline}
\label{Pxs1}
P(x_s)=  C_s^2\iint dx\rq dx\rq{}\rq\braket {\Psi} {x\rq} e^{-\frac{(x'-x'')^2}{4\sigma_w^2}}  \braket {x\rq{}\rq} {\Psi} \times \\
\times\left(\int dxe^{-\frac{(x_s-x) ^2}{\sigma_s^2}}\bra{x\rq}\Ud\ket{x}\bra{x}\U\ket{x\rq{}\rq}\right).
\end{multline}
For a particle of mass $m$ that evolves under a flat potential during $\tau$, we can evaluate $\bra x \U \ket{x\rq}$ using \cite{shankar}
\begin{equation}
\label{evo}
\bra x \U \ket{x\rq}=\left( i\pi(2\hbar \tau/m)\right)^{-1/2} e^{\frac{i(x-x\rq)  ^{2}}{(2\hbar \tau/m)}}.
\end{equation}
Substituting \eref{evo} into \pref{Pxs1} and solving the integral between parenthesis, we have
\begin{multline}
\label{Pxs2}
P(x_s)=  \iint dx\rq dx\rq{}\rq e^{-\frac{(x'-x'')^2}{4\sigma_w^2}}   e^{-\left(\frac{\sigma_s m}{2\hbar \tau}\right)^2(x\rq-x\rq{}\rq) ^2} \times \\
\braket {\Psi} {x\rq} \bra{x\rq}\Ud\ket{x_s}\bra{x_s}\U\ket{x\rq{}\rq} \braket {x\rq{}\rq} {\Psi}.
\end{multline}
One easily realizes that the probability in \pref{Pxs2} can be computed as $P(x_s)=\bra\Psi\Ud\Sad\Sa\Ud\ket\Psi$ when the following limit is satisfied,
\begin{equation}
\label{lim}
\frac {\sigma_w} {\tau} \gg  \frac{\hbar}{m\sigma_{s}}.
\end{equation}
Let us emphasize that this condition, includes Wiseman\rq{}s result \cite{Wiseman} as a particular case: $\sigma_w \rightarrow \infty$, $\sigma_s \rightarrow 0$ and $\tau \rightarrow 0$. Our development will justify the effective measurement of the Bohmian velocity (up to a negligible error) for a broad range of $\sigma_w$, $\sigma_s$ and $\tau$.

Identical steps can be done for the evaluation of ${\int dx_w  x_w P(x_w\cap x_s)}$ in  \eref{cond}. The only difference resides on the integration on $x_w$, which in this case gives $(x\rq{}+x\rq{}\rq{})/2\exp[-(x'-x'')^2/4\sigma_w^2]$.
Using $\int dx \hspace{0.02in}x \ket {x}\bra {x} =\x$, under the limit \pref{lim}, we obtain ${\int dx_w  x_w P(x_w\cap x_s)}={\Real(\bra\Psi\Ud\Sad\Sa \U\x\ket\Psi )}$ . Finally, we can rewrite \eref{cond} as:
\begin{equation}
\label{AAV1}
\E[x_w|x_s]=\textstyle\frac{\Real(\bra\Psi\Ud\Sad\Sa\U\x\ket\Psi )}{\bra\Psi\Ud\Sad\Sa\U\ket\Psi} .
\end{equation}
Next, we define the following (averaged) position $\bar x_s=\bra\Psi\Ud\Sad\Sa\x\U\ket\Psi/\bra\Psi\Ud\Sad\Sa\U\ket\Psi$, so that using \eref{AAV1} and the commutator $[\U,\x]$, we get:
\begin{equation}
\label{AAV2}
\bar x_s-\E[x_w|x_s]=\textstyle\frac {\Real(\bra\Psi\Ud\Sad\Sa [\U,\x]\ket\Psi )}{\bra\Psi\Ud\Sad\Sa\U\ket\Psi},
\end{equation}
without any reference to $\Wa$.
To further develop \eref{AAV2}, we evaluate the commutator $[\U,\x]$ using the Maclaurin series for $\U$:
\begin{equation}
\label{maclaurin}
[\U,\x]  =\textstyle\sum\limits_{n=1}^{\infty}\frac{(-i)  ^n\tau^n}{n!\hbar^n}[\Ha^n ,\x],
\end{equation}
where $\Ha=\p^2/2m+V$ is the system Hamiltonian with $V$ a flat potential at the spatial region where the wave function is different from zero during the time between measurements. No restriction on $V$ for other regions and times. Given two operators $\hat A$ and $\hat B$, it can be proven that $[\hat A^n,\hat B]=\textstyle\sum_{j=1}^{n}\hat A^{j-1}[\hat A,\hat B] \hat A^{n-j}$. Then, being $[\Ha,\x]=-i\hbar/m\p$ and $[\Ha,\p]=0$, the commutator $[\Ha^n,\x]$ gives:
\begin{equation}
\label{Hanx}
[\Ha^n,\x] =-\textstyle\frac{i\hbar n}{m}\p\Ha^{n-1},
\end{equation}
and substituting \eref{Hanx} into \eref{maclaurin} we obtain:
\begin{equation}
\label{commutator}
[\U,\x]=-{\frac{\tau}{m}}\p\U,
\end{equation}
without considering the limit $\tau\rightarrow0$.
Using \eref{commutator} and the definition \pref{S}, a straightforward calculation for the numerator of \eref{AAV2} gives:
\begin{eqnarray}
\Real(\bra\Psi\Ud\Sad\Sa[\U,\x]\ket\Psi ) \equiv \tau \bar J(x_s,t_s) =\nonumber\\
\tau C_s^2 \int dx J(x,t_s)\exp[-(x_s-x)^2/\sigma_s^2],
\label{nolabel}
\end{eqnarray}
where $J(x,t_s)$ is the standard quantum current probability density \cite{libro}. Similarly, we define ${\bra\Psi\Ud\Sad\Sa\U\ket\Psi} = C_s^2 \int dx |\Psi(x,t_s)|^2\exp[-(x_s-x)^2/\sigma_s^2]\equiv |\bar \Psi(x_s,t_s)|^2$ for the denominator. Finally, the velocity, defined as \eref{AAV2} divided by $\tau$, gives:
\begin{equation}
\label{velobohm_s}
\bar v(x_s,t_s)=\frac {\bar x_s - E[x_w|x_s]} {\tau}=\frac {\bar J(x_s,t_s)} { |\bar{\Psi}(x_s,t_s)|^2}.
\end{equation}
This expression is just the Gaussian-spatially-averaged current density $\bar J(x_s,t_s)$ inside a tube of diameter $\sigma_s$ divided by the corresponding Gaussian-spatially-averaged probability $|\bar \Psi(x_s,t_s)|^2$.

Whether or not the Gaussian-spatially-averaged value \pref{velobohm_s} is identical to the Bohmian velocity depends on the measuring apparatus resolution, i.e. $\sigma_s$, and the de Broglie wavelength $\lambda$ associated to $\ket \Psi$. Under the limit
\begin{equation}
\label{lim2}
\sigma_s<  \lambda,
\end{equation}
one can assume $\Psi(x,\tau) \approx \Psi(x_s,t_s)$ for $x\in [x_s-\sigma_s,x_s+\sigma_s]$, so that $\bar \Psi(x_s,t_s)\approx \Psi(x_s,t_s)$. Identically, $\bar J(x_s,t_s) \approx J(x_s,t_s)$ and $\bar x_s=x_s$. Then, \eref{velobohm_s} directly recovers the Bohmian velocity  $\bar v(x_s,t_s) \approx v$ with:
\begin{equation}
\label{velobohm}
v \equiv v(x_s,t_s) =\frac {J(x_s,t_s)}{|\Psi(x_s,t_s)|^2}.
\end{equation}
Let us mention that the consideration $\sigma_s  \approx \lambda$ and the momentum $p=h/\lambda$ implies $\hbar/(m\sigma_s) \approx v$ in the limit \pref{lim}.

>From the definition of velocity in \pref{vexp}, one could reasonably expect to get a value associated to the velocity \emph{averaged} during the time interval $\tau$ and associated to a \emph{perturbed} wave function. However, under the conditions \pref{lim} and \pref{lim2}, the result \pref{velobohm} is clearly identified as the \emph{instantaneous} (bohmian) velocity  associated with an \emph{unperturbed} wave function at the final time $t_s$. The mathematical reasons leading to \pref{velobohm} are fully detailed in the previous calculations. Here, we try to provide some physical insights. It is well known that a measurement process induces a perturbation on the wave function, breaking the symmetry in its time evolution. In our case,  because of the imposed conditions \pref{lim} and \pref{lim2}, the roles of the first and second measurements are very different. The condition \pref{lim} implies that the first measurement perturbs very weakly the wave function in a neighborhood $I_w$ of radius $\sigma_w$ around $x_w$, while the second limit \pref{lim2} implies a very strong perturbation of the wave function during the second measurement process. As a result, when constructing \pref{vexp}, only the position eigenstates belonging to $I_w$ (where the wave function remains mainly \emph{unperturbed} by the first measurement)  are used. In fact, the ensemble average \pref{AAV1} has no memory of the first measurement process (i.e., of the first POVM). Moreover, the condition of flat potential between the two measurements that leads to Eq. \pref{commutator} implies explicit independence of $\tau$ because it provides \emph{free} evolution of the \emph{unperturbed} wave function. In this regard, the first measurement does not actually break the symmetry. The obvious consequence (supported by our calculation) is that the velocity in \pref{vexp} is independent of the time $\tau$ between the two measurements. Finally, since the symmetry is broken essentially by the second measurement, the velocity that we obtain is the one associated with an \emph{unperturbed} wave function at the last time $t_s$.

Another way of explaining our results is by noticing that the identity \pref{commutator} can be used for a finite $\tau$  because we assume that the potential is flat at the spatial region where the wave function is different from zero. For a classical system evolving under a flat potential from $t_w$ till $t_s=t_w+\tau$, the instantaneous velocity at $t_s$ is exactly equal to the averaged velocity during $\tau$. The classical velocity remains constant during this time interval because the classical acceleration is zero. In the quantum counterpart, from Ehrenfest theorem, we know that the ensemble momentum with a flat potential is constant during $t_w< t\le t_s$. Using the limit \pref{lim}, the ensemble momentum can be defined as $\bra{\Psi(t)}\p \ket{\Psi(t)}=\int\bra{\Psi(t_w)}\Wad U_{t-t_w}^\dagger\p U_{t-t_w}\Wa\ket{\Psi(t_w)}dx_w$ which corresponds to \pref{nolabel} without performing the second measurement. This again justifies why the resulting velocity evaluated with our protocol is independent of $\tau$ and exactly equal to the (Bohmian) velocity measured at $t_s$.


\subsection{Calculation of the ensemble velocity variance}

Let us now compute the velocity variance. Since  $x_s$ and $\tau$ are fixed in \eref{vexp}, $var(v_{e})=var(x_w)/\tau^2$. Thus,  $var(x_w)=E[x_w^2|x_s]-(E[x_w|x_s])^2$ where $E[x_w|x_s]$ defined in \eref{cond} is obtained from  \eref{velobohm}. The evaluation of ${\int dx_w  x_w^2 P(x_w\cap x_s)}$ follows identical steps as in the computation of $P(x_s)$, where again the only difference resides in the integral in $x_w$ that now gives $( \sigma_w^2/2+(x\rq{}+x\rq{}\rq{})^2/4) \exp[-(x\rq{}-x\rq{}\rq{})^2/4 \sigma_w^2]$.
Using again $\int dx \hspace{0.02in}x  \ket {x}\bra {x} =\x$ and $\int dx \hspace{0.02in}x^2  \ket {x}\bra {x} =\x^2$, the final result, under the limit \pref{lim}, is:
\begin{multline}
\label{cond4}
\E[x_w^2 | x_s] =\textstyle\frac 1 2 \sigma_w^2+\frac 1 2 \frac{\Real(\bra\Psi\Ud\Sad\Sa\U\x^2\ket\Psi )}{\bra\Psi\Ud\Sad\Sa\U\ket\Psi}\\
+ \textstyle\frac 1 2 \frac{\Real(\bra\Psi\x\Ud\Sad\Sa\U\x\ket\Psi )}{\bra\Psi\Ud\Sad\Sa\U\ket\Psi},
\end{multline}
which, as detailed in \sref{appendixA} in the Appendix, finally gives
\begin{equation}
\label{var}
var(v)= \frac {\sigma_w^2} {2\tau^2}+\textstyle\frac 2 m Q_B(x_s)+O\left(\frac{\hbar}{m\tau}\right),
\end{equation}
where $Q_B(x_s)$ is the (local) Bohmian quantum potential \cite{Bohm, libro}. Under the limits \pref{lim} and \pref{lim2}, the term ${\sigma_w^2} /(2{\tau^2})$ in Eq.~(\ref{var}) will be orders of magnitude greater than the other two. For an experimentalist, this means that the presence of the quantum potential on the spatial fluctuations of \eref{var} will be hardly accessible, and that $var(v)$ provides basically the value $\sigma_w$ of the apparatus. Using the well know result from the probability calculus $\varepsilon(N)={\sqrt{var(v)}}/{\sqrt{N}}\approx {\sigma_w}/({\tau\sqrt{2N}})$, such variance can be used to evaluate the number $N$ of measurements needed to obtain \pref{velobohm} with a given error $\varepsilon(N)$.


\subsection{Error analysis}

In order to test how robust (i.e. how independent of $\sigma_w$, $\sigma_s$ and $\tau$) is the possibility of measuring the Bohmian velocity in a laboratory, we compute the (local) error $\varepsilon_w(x_s) \equiv  \abs{v_{e}(x_s)-\bar {v}(x_s)}$. The details of the calculation are reported in \sref{appendixB} in the Appendix:
\begin{equation}
\label{errorW}
\varepsilon_w(x_s)=\frac{\tau\hbar^2}{4m^2\sigma_w^2}\abs{
\dfrac{2(  1-\tau\partial_x v)\partial_x \rho-\tau\rho\partial_x^2 v} {\rho+\frac{\tau^{2}\hbar^{2}%
}{4m^{2}\sigma_w^{2}}\partial_x^{2}\rho}},
\end{equation}
where $\rho=\abs{\psi(x_s,t_s)}^2$. We further define the measuring apparatus error $\varepsilon_{s}(x_s) \equiv \abs{v(x_s)-\bar {v}(x_s)}$ deriving from the requirement \pref{lim2}. The calculation reported in \sref{appendixC} in the Appendix gives:
\begin{equation}
\label{errorS}
\varepsilon_{s}(x_s) =\sigma_s^2\abs{ \frac{\frac{2}{\tau}\partial_x\rho+(  2\partial_x\rho-\rho \partial_x )  \partial_x v}{4\rho+\sigma_s^2\partial_x^2\rho}}.
\end{equation}
It is worth noticing that, by construction, the total error $\varepsilon(x_s) \equiv \abs{v(x_s)-v_e(x_s)}$ accomplishes $\varepsilon(x_s) \leq \varepsilon_s(x_s)+\varepsilon_w(x_s)$.


\section{Ensemble current density}

We observe that the same set of measured values $x_w$ and $x_s$ can be used to define an experimental current density: 
\begin{equation}
\label{encurrentg}
J_e(x_s,t_s) =  \frac  {P(x_s)x_s-\int dx_wx_wP(x_w\cap x_s)} {\tau}.
\end{equation}
The get experimental value $J_e(x_s,t_s)$, we do only need to change how the measured data $x_w$ and $x_s$ is treated. The fact that expression \pref{encurrentg} provides the expected theoretical definition of the current density (within a negligible error) can be straightforwardly computed following previous developments of ${P(x_s)}$ and ${\int dx_w  x_w P(x_w\cap x_s)}$ in \sref{sec_velocity}. Identically, all the previous calculations for the variance of the current density and their errors can be then repeated for the current in a similar way.


\begin{figure}
\centerline{
\includegraphics[width=0.97\columnwidth]{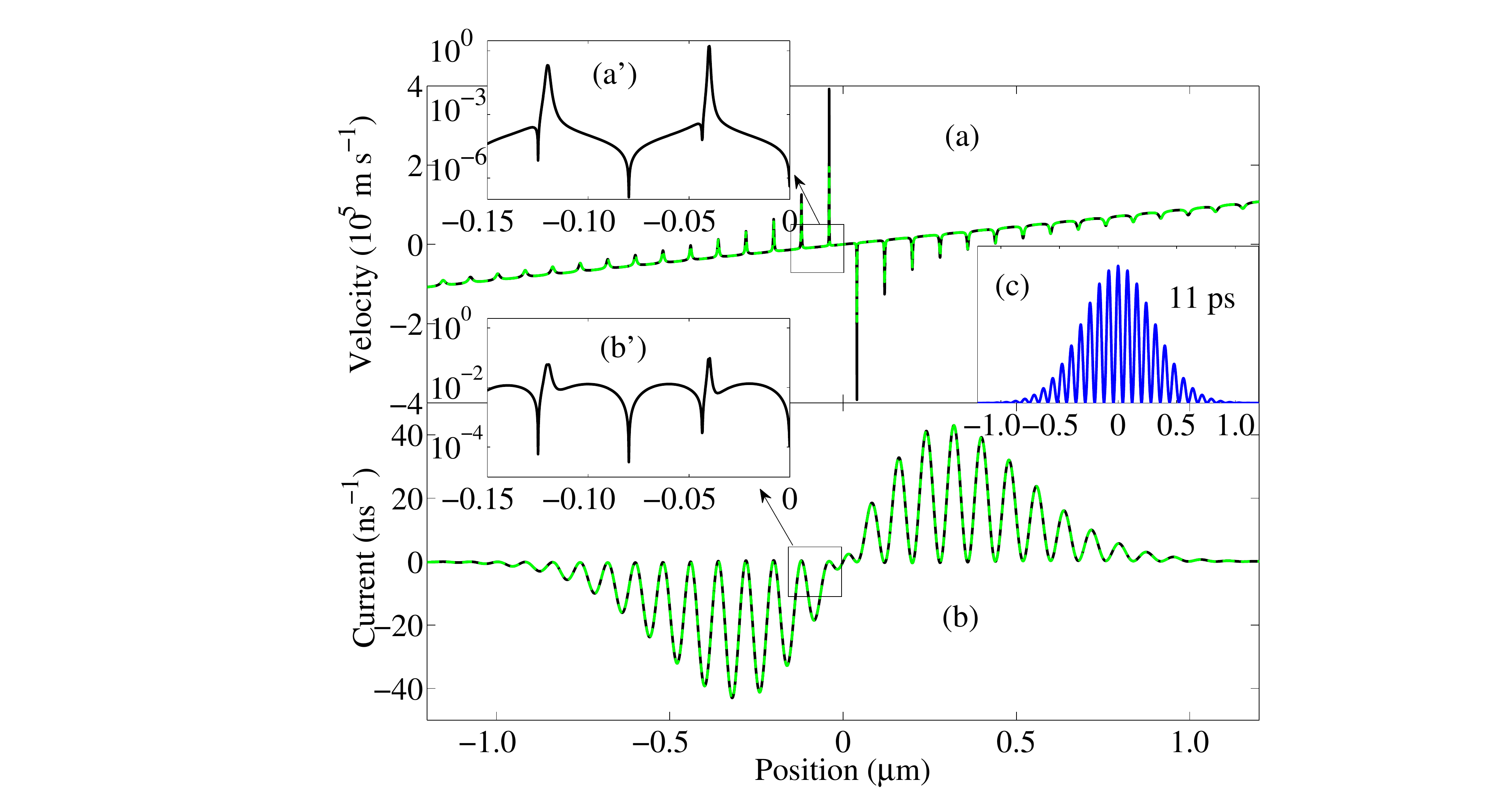}}
\caption{\label{vel} (Color online) (a) Velocity distribution ($v$ black solid line and $v_e$ green dashed line) and (b) quantum current density ($J$ black solid line and $J_e$ green dashed line) for an electron in a double slit experiment at $t_s=11$ps,  $\sigma_s=0.2$nm and $\sigma_w=150$nm. The two insets (a\rq{}) and (b\rq{}) are the total error $\varepsilon_s(x_s)+\varepsilon_w(x_s)$ in the highlighted position interval for the velocity and current, respectively. Inset (c) is $|\Psi|^2$ at $t_s=11$ps.}
\end{figure}

\section{Numerical results and discussion}

As a numerical test of our prediction, we consider an electron passing through a double slit. For simplicity, the time evolution of two 1D initial Gaussian wave-packets with zero central momenta and central positions separated a distance of 100 nm are explicitly simulated. This roughly corresponds to the evolution of the quantum state after crossing the double-slit at $t=0$s. From \fref{vel}(a) the agreement between the exact Bohmian velocity $v$ in \pref{velobohm} and $v_e$ [numerically evaluated from \pref{vexp}, \pref{cond}, \pref{prob1}  and \pref{prob2} without any limit or approximation] is excellent and it is highlighted by the inset \ref{vel}(a\rq{}) where the total error \pref{errorW} plus \pref{errorS} is reported.
\begin{figure}
\centerline{
\includegraphics[width=0.97\columnwidth]{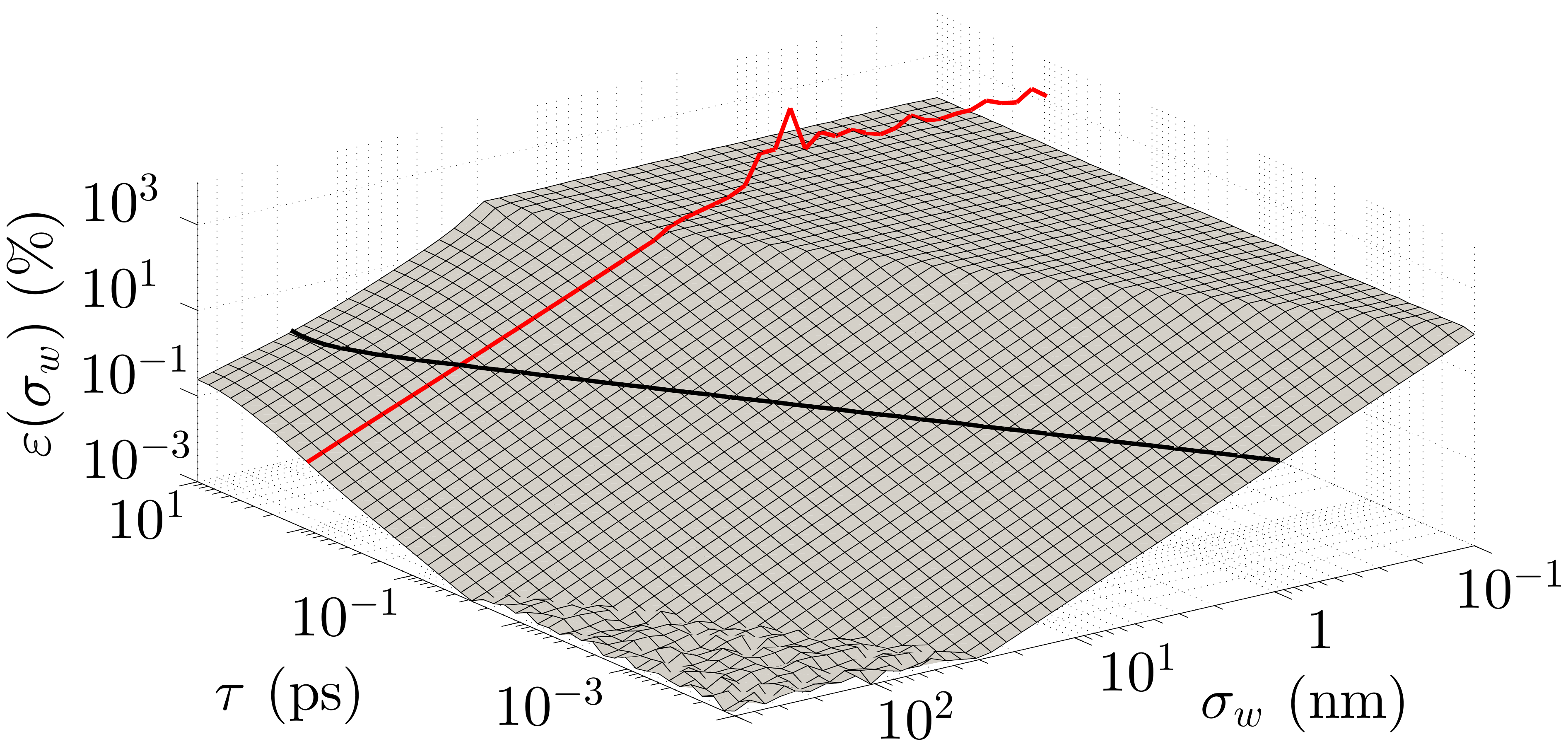}}
\caption{\label{tau_sigma_w} (Color online) Relative error $\varepsilon_w$ integrated over all positions $x_s$ as a function of $\sigma_w$ and $\tau$ for $\sigma_s=0.2$nm for the numerical test represented in \fref{vel}. Black line bounds the region for $\varepsilon(\sigma_w)\le1\%$ and red line is the analytical error for the value $\tau=1$ps.}
\end{figure}
\begin{figure}
\centerline{
\includegraphics[width=0.97\columnwidth]{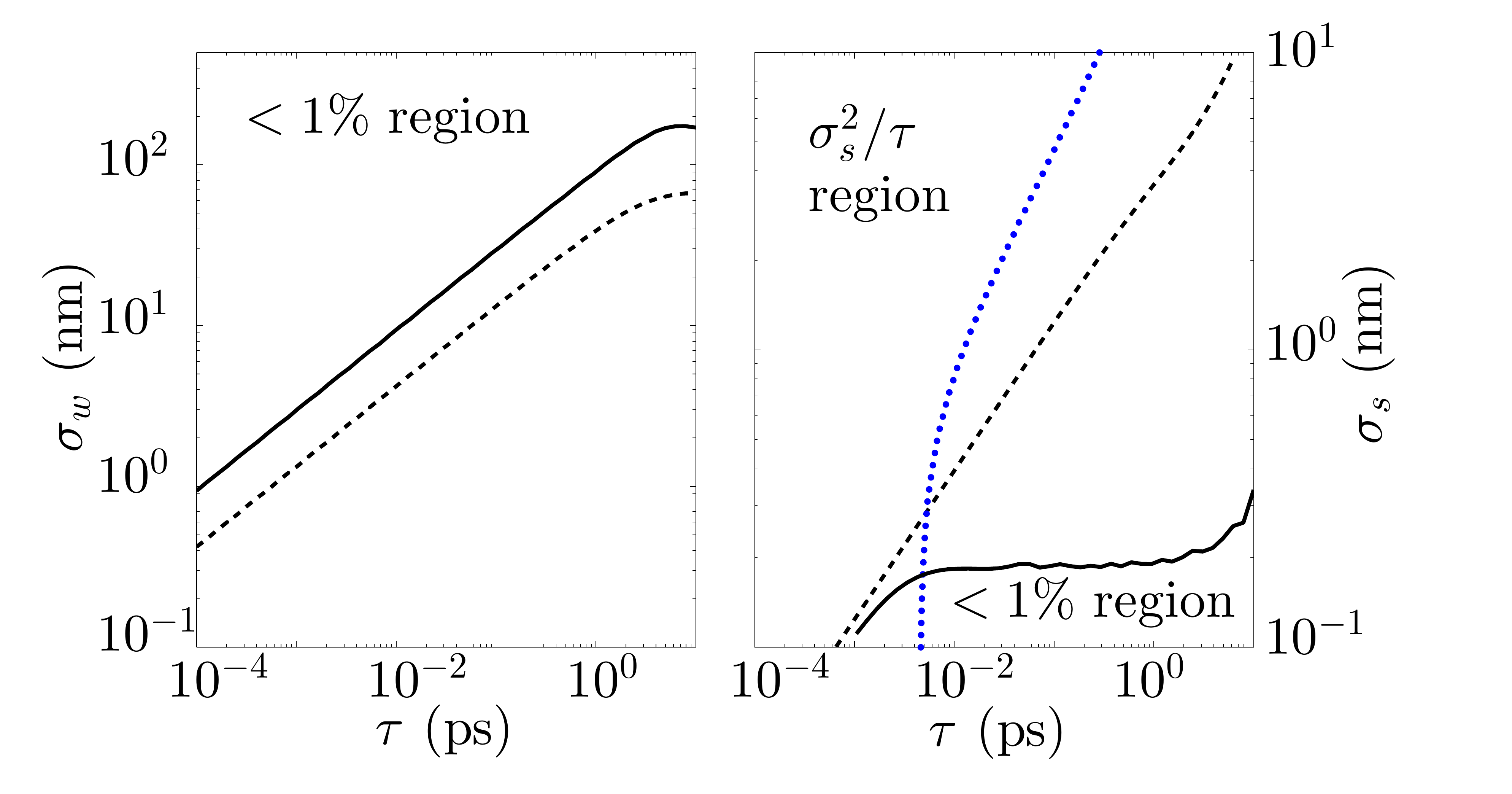}}
\caption{\label{tau_sigma_s} (Color online) Left inset, region of relative error $\varepsilon_w<1\%$ and right inset region of relative error $\varepsilon_s<1\%$. Solid lines are the boundaries for the velocity and dashed line are the boundaries for the quantum current. Dotted line bounds the $\sigma_s^2/\tau$ region.}
\end{figure}
In  \fref{tau_sigma_w}, we plot the normalized value of the error $\varepsilon_w(x_s)$ integrated over $x_s$ as $\varepsilon_w=(\int dx_s \varepsilon_w(x_s)^2/\int dx_s v(x_s)^2)^{1/2}$. The main conclusion extracted from \fref{tau_sigma_w} is that a large set of parameters (large $\sigma_w/\tau$ values) allows a very accurate  measurement of the Bohmian velocity, justifying the robustness of our proposal.


At this point, we emphasize some relevant issues. First, we have shown theoretically and  numerically that the Bohmian velocity of an unperturbed state under general laboratory conditions can be obtained from two POVM measurements separated by a finite $\tau$. Unlike the results derived from the AAV formulation \cite{AAV}, the limits \pref{lim} and \pref{lim2} provide a simple quantitative explanation of the experimental conditions for an accurate and robust  measurement of the Bohmian velocity.

On the other hand, the error $\varepsilon_s(x_s)$ in \pref{errorS} has a term that diverges as $\sigma_s^2/\tau$, meaning that a $\tau$ close to zero will produce an inaccurate measurement of the velocity for finite $\sigma_s$. This regime is reported in the right inset of \fref{tau_sigma_s}. Roughly speaking, for $\tau \rightarrow 0$, the wave packet moves a distance $v \tau$. When $v \tau < \sigma_s$ the measured position $x_s$ has no relation to the velocity. We emphasize again that Wiseman's result \cite{Wiseman} does not suffer from this inaccuracy because he considers, both, $\sigma_s \rightarrow 0$ and $\tau \rightarrow 0$.

A closer look at the expressions \pref{errorW} and \pref{errorS} shows that the error diverges when $\rho$ has oscillations with minima tending to zero. This can be clearly seen in \fref{vel}(a) and (a\rq{}) where the highest peak of the velocity corresponds to a minimum of $\rho$ very close to zero. This situation is reversed when we evaluate the current $J$ [see \fref{vel}(b) and (b\rq{})]. In fact, in these critical points, $J\rightarrow 0$ and even the corresponding errors become very small. In \fref{tau_sigma_s} it is evident the shift of the $<1\%$ region due to this error reduction.

Perhaps, the most surprising feature of our protocol is that a local (in time and position) Bohmian velocity can be measured with a large temporal separation between measurements, while one would expect a time-averaged value as discussed at the end of section \ref{sec_velocity}.
This is highly counter-intuitive because we are in a scenario where the time-evolving interferences implies large acceleration of the Bohmian particle in order to rapidly avoid the nodes of the wave function.

Finally, another relevant result is that the accuracy of the Bohmian velocity is obtained at the prize of increasing the dispersion on $x_w$ (as seen in \eref{var} for large $\sigma_w$).  Therefore, the fact that we can obtain the Bohmian velocity is not because the system remains unperturbed after one position measurement, rather because of the ability of the ensemble average done in the  $x_w$ integrals on \eref{prob1} and \eref{prob2} to compensate for the different perturbations. The fact that a very large perturbation of the state is fully compatible with a negligible error can be easily seen in our numerical data. The {\it measured} state is roughly equal to the product of the unperturbed wave function (whose support is $L\approx 2000 $nm at time $t_w=11$ps in \fref{vel}) by a Gaussian function centered at the measured position with a dispersion equal to $\sigma_w$ (for example, $\sigma_w \approx 150 $nm for $\tau=1$ps in \fref{tau_sigma_w}). Even for $\sigma_w << L$ (i.e. a large perturbation), the velocity error is negligible in \fref{tau_sigma_w}.

\section{Conclusions}

The work presented here explains a protocol for measuring Bohmian velocities. It is based on using an ensemble of two position measurements separated by a finite time interval. The perturbation of each position measurements on the state is modeled by a POVM. The difference  between the Bohmian velocity of the unperturbed state and the ensemble  Bohmian velocity of the two-times measured state is predicted to be much smaller than $1 \%$ in a large range of parameters. The work clarifies the laboratory conditions necessary for measuring Bohmian velocities, while relaxing the experimental setup by allowing reasonable position uncertainties and a finite time interval between measurements. Following the same ideas presented in this work (with two POVM for position measurements) an equivalent analysis for the case of combined POVM momentum plus POVM position measurements can be carried out for particles with mass. This case, experimentally tested also for relativistic photons \cite{Bohmphoton}, could be of major interest for several experiments. In this sense, a clear and feasible proposal has been recently presented for the demonstration of the nonlocal character of Bohmian mechanics by measuring the ensemble velocities of path-entangled particles \cite{nou}. Finally, as mentioned in the introduction, the present work is fully developed within orthodox quantum mechanics. However, we emphasize that this works opens relevant and unexplored possibilities for understanding quantum phenomena through the quantitative comparison between simulated and measured Bohmian (or hydrodynamic) trajectories \cite{libro,Oriols,turbulence}, instead of using the wave function and its related parameters.

\section*{acknowledgement}

We want to acknowledge S. Goldstein, D. D\"{u}rr, N.Vona, J. Mompart, A.Benseny and A. Segura for insightful discussions. This work has been partially supported by the Spanish government through projects TEC2012-31330 and TEC2011-14253-E and from the US/DOE grant DE-FG02-05ER46204 and through the Beatriu de Pinos project 2010 BP-A 00069.

\appendix
\numberwithin{equation}{section}

\section{Derivation of the variance}
\label{appendixA}

In order to evaluate the variance $var(v)=var(x_{w}^{2})$ defined as
\[
var(x_{w}^{2})=\frac{\int dx_{w}x_{w}^{2}P(x_{w}\cap x_{s})}{P(x_{s}%
)}-(E[x_{w}|x_{s}])^{2},
\]
where $P(x_{w}\cap x_{s})$ and $P(x_{s})$ are given respectively by Eq. \pref{prob1} and \pref{prob2}, we calculate
\begin{multline}
\int dx_{w}x_{w}^{2}P(x_{w}\cap x_{s})=\dfrac{\sigma_{w}^{2}}{2}\langle
\Psi|U_{\tau}^{\dag}\hat{S}_{s}^{\dag}\hat{S}_{s}U_{\tau}|\Psi\rangle+\\
+C_{s}^{2}\iiint dxdx^{\prime}dx^{\prime\prime}\left(  \frac{x^{\prime
}+x^{\prime\prime}}{2}\right)  ^{2}\times \\
\times e^{-\frac{\left(  x^{\prime}-x^{\prime
\prime}\right)  ^{2}}{4\sigma_{w}^{2}}}e^{-\frac{\left(  x_{s}-x\right)  ^{2}%
}{\sigma_{s}^{2}}}|x^{\prime}\rangle\langle x^{\prime}|U^{\dag}|x\rangle
\langle x|U|x^{\prime\prime}\rangle\langle x^{\prime\prime}|,
\end{multline}
where the integral over $x_{w}$ has been already evaluated. From Eq. \pref{evo} and
accounting for the limit \pref{lim} we have%
\begin{multline}
\int dx_{w}x_{w}^{2}P(x_{w}\cap x_{s})=\dfrac{\sigma_{w}^{2}}{2}\langle
\Psi|U_{\tau}^{\dag}\hat{S}_{s}^{\dag}\hat{S}_{s}U_{\tau}|\Psi\rangle+\\
+\tfrac{1}{2}\operatorname{Re}(\langle\Psi|U_{\tau}^{\dag}\hat{S}_{s}^{\dag
}\hat{S}_{s}U_{\tau}\hat{x}^{2}|\Psi\rangle)+\tfrac{1}{2}\langle\Psi|\hat
{x}U_{\tau}^{\dag}\hat{S}_{s}^{\dag}\hat{S}_{s}U_{\tau}\hat{x}|\Psi
\rangle.\label{x2}%
\end{multline}
Under the limit \pref{lim} we have shown in the text that $P(x_{s})=\langle
\Psi|U_{\tau}^{\dag}\hat{S}_{s}^{\dag}\hat{S}_{s}U_{\tau}|\Psi\rangle$. Moreover using Eq. (16) we have
\begin{multline}
\langle\Psi|\hat{x}U_{\tau}^{\dag}\hat{S}_{s}^{\dag}\hat{S}_{s}U_{\tau}\hat
{x}|\Psi\rangle=\operatorname{Re}(\langle\Psi|U_{\tau}^{\dag}\hat{S}_{s}%
^{\dag}\hat{S}_{s}U_{\tau}\hat{x}^{2}|\Psi\rangle+\\
+\dfrac{\tau}{m}\langle
\Psi|U_{\tau}^{\dag}[\hat{S}_{s}^{\dag}\hat{S}_{s},\hat{p}]U_{\tau}\hat
{x}|\Psi\rangle),
\end{multline}
that substituted in Eq. \pref{cond4} gives
\begin{multline}
var(x_{w}^{2})=\dfrac{\sigma_{w}^{2}}{2}+\frac{\operatorname{Re}(\langle
\Psi|U_{\tau}^{\dag}\hat{S}_{s}^{\dag}\hat{S}_{s}U_{\tau}\hat{x}^{2}%
|\Psi\rangle)}{\langle\Psi|U_{\tau}^{\dag}\hat{S}_{s}^{\dag}\hat{S}_{s}%
U_{\tau}|\Psi\rangle}+\\
+\frac{\tau}{2m}\frac{\operatorname{Re}(\langle
\Psi|U_{\tau}^{\dag}[\hat{S}_{s}^{\dag}\hat{S}_{s},\hat{p}]U_{\tau}\hat
{x}|\Psi\rangle)}{\langle\Psi|U_{\tau}^{\dag}\hat{S}_{s}^{\dag}\hat{S}%
_{s}U_{\tau}|\Psi\rangle}-(E[x_{w}|x_{s}])^{2}.\label{var1}%
\end{multline}
The difference between the second and the fourth terms on the r.h.s. of Eq.
(\ref{var1}) can be rewritten using again Eq. \pref{AAV1} and \pref{commutator} as
\begin{multline}
\frac{\operatorname{Re}(\langle\Psi|U_{\tau}^{\dag}\hat{S}_{s}^{\dag}\hat
{S}_{s}U_{\tau}\hat{x}^{2}|\Psi\rangle)}{\langle\Psi|U_{\tau}^{\dag}\hat
{S}_{s}^{\dag}\hat{S}_{s}U_{\tau}|\Psi\rangle}-(E[x_{w}|x_{s}])^{2}%
=\tfrac{\tau^{2}}{m^{2}}\times\\
\times \left(  \dfrac{\operatorname{Re}\langle\Psi|U_{\tau
}^{\dag}\hat{S}_{s}^{\dag}\hat{S}_{s}\hat{p}^{2}U_{\tau}|\Psi\rangle}%
{\langle\Psi|U_{\tau}^{\dag}\hat{S}_{s}^{\dag}\hat{S}_{s}U_{\tau}|\Psi\rangle
}-\left(  \dfrac{\operatorname{Re}\langle\Psi|U_{\tau}^{\dag}\hat{S}_{s}%
^{\dag}\hat{S}_{s}\hat{p}U_{\tau}|\Psi\rangle}{\langle\Psi|U_{\tau}^{\dag}%
\hat{S}_{s}^{\dag}\hat{S}_{s}U_{\tau}|\Psi\rangle}\right)  ^{2}\right)
.\label{var2}%
\end{multline}
Using in \pref{var2} the relations $\langle x|\hat{p}U_{\tau}|\Psi\rangle=-i\hbar\partial
_{x}\Psi(x,\tau)$ and $\langle x|\hat{p}^{2}U_{\tau}|\Psi\rangle=-\hbar
^{2}\partial_{x}^{2}\Psi(x,\tau)$ and the limit \pref{lim2}, we can rewrite (\ref{var2}) as:
\begin{multline}
var(x_{w}^{2})=\dfrac{\sigma_{w}^{2}}{2}+2\dfrac{\tau^{2}}{m}Q_{B}(x_{s}%
,\tau)+\\
+\frac{\tau}{2m}\frac{\operatorname{Re}(\langle\Psi|U_{\tau}^{\dag}%
[\hat{S}_{s}^{\dag}\hat{S}_{s},\hat{p}]U_{\tau}\hat{x}|\Psi\rangle)}%
{\langle\Psi|U_{\tau}^{\dag}\hat{S}_{s}^{\dag}\hat{S}_{s}U_{\tau}|\Psi\rangle
}.\label{var3}%
\end{multline}
We further evaluate the commutator $[\hat{S}_{s}^{\dag}\hat{S}_{s},\hat{p}]$
as
\begin{multline}
\lbrack\hat{S}_{s}^{\dag}\hat{S}_{s},\hat{p}]|\Psi\rangle=-i\hbar C_{s}%
^{2}\int dx\left(  e^{-\frac{\left(  x_{s}-x\right)  }{\sigma_{s}^{2}}^{2}%
}\left(  \partial_{x}\Psi(x)\right)  |x\rangle-\right. \\
-\left.\left[  \partial_{x}\left(
e^{-\frac{\left(  x_{s}-x\right)  }{\sigma_{s}^{2}}^{2}}\Psi(x)\right)
\right]  |x\rangle\right)  =-i\hbar\partial_{x_{s}}\left(  \hat{S}_{s}^{\dag
}\hat{S}_{s}\right)  |\Psi\rangle,\label{commutator_appendix}%
\end{multline}
and using Eq. (\ref{commutator_appendix}) in the last term of Eq. (\ref{var3}) we have
\begin{multline}
var(x_{w}^{2})=\dfrac{\sigma_{w}^{2}}{2}+2\dfrac{\tau^{2}}{m}Q_{B}(x_{s}%
,\tau)+\\
+\frac{\tau\hbar}{2m}\frac{\partial_{x_{s}}\operatorname{Im}(\langle
\Psi|U_{\tau}^{\dag}\hat{S}_{s}^{\dag}\hat{S}_{s}U_{\tau}\hat{x}|\Psi\rangle
)}{\langle\Psi|U_{\tau}^{\dag}\hat{S}_{s}^{\dag}\hat{S}_{s}U_{\tau}%
|\Psi\rangle}.\label{var4}%
\end{multline}
>From the limits \pref{lim} and \pref{lim2} we have
\begin{equation}
\frac{\tau\hbar}{m}\ll\sigma_{w}\sigma_{s}\ll\sigma_{w}^{2},%
\end{equation}
and we can conclude that both the last two terms of the r.h.s. of Eq. (\ref{var4})
are much smaller than $\sigma_{w}^{2}$.

\section{Derivation of the error $\varepsilon_{s}(x_{s})$}
\label{appendixB}

The definition of $\varepsilon_{s}(x_{s})$ is:
\begin{multline}
\varepsilon_{s}(x_{s})=|v(x_{s})-\bar{v}(x_{s})|=\tau^{-1}\left\vert
\dfrac{\operatorname{Re}\langle\Psi|U_{\tau}^{\dag}\hat{S}_{s}^{\dag}\hat
{S}_{s}U_{\tau}\hat{x}|\Psi\rangle}{\langle\Psi|U_{\tau}^{\dag}\hat{S}%
_{s}^{\dag}\hat{S}_{s}U_{\tau}|\Psi\rangle}-\right.\\
\left.-\dfrac{\operatorname{Re}%
\langle\Psi|U_{\tau}^{\dag}|x_{s}\rangle\langle x_{s}|U_{\tau}\hat{x}%
|\Psi\rangle}{\langle\Psi|U_{\tau}^{\dag}|x_{s}\rangle\langle x_{s}|U_{\tau
}|\Psi\rangle}\right\vert .\label{epss}%
\end{multline}
We can easily take the limit of (\ref{epss}) for $\sigma_{s}$ small using a Taylor
series,
\begin{multline}
\langle\Psi|U_{\tau}^{\dag}\hat{S}_{s}^{\dag}\hat{S}_{s}U_{\tau}|\Psi\rangle=\\%
={\textstyle\sum_{n=0}^{2}}
\frac{\partial_{x}^{n}\rho}{n!}C_{s}^{2}\int e^{-\frac{\left(  x_{s}-x\right)
}{\sigma_{s}^{2}}^{2}}\left(  x-x_{s}\right)  ^{n}dx=\\
=\rho+\frac{\sigma_{s}%
^{2}}{4}\partial_{x}^{2}\rho\label{1}%
\end{multline}
and in the same way using Eq. \pref{commutator}
\begin{multline}
\operatorname{Re}\langle\Psi|U_{\tau}^{\dag}\hat{S}_{s}^{\dag}\hat{S}%
_{s}U_{\tau}\hat x|\Psi\rangle=\\
=\operatorname{Re}\langle\Psi|U_{\tau}^{\dag}\hat
{S}_{s}^{\dag}\hat{S}_{s}\hat x U_{\tau}|\Psi\rangle-\frac{\tau}{m}\operatorname{Re}%
\langle\Psi|U_{\tau}^{\dag}\hat{S}_{s}^{\dag}\hat{S}_{s}\hat p U_{\tau}|\Psi
\rangle=\label{2}\\
x_{s}\rho+\frac{\sigma_{s}^{2}}{2}\partial_{x}\rho+x_{s}\frac{\sigma_{s}}%
{4}\partial_{x}^{2}\rho-\tau J-\tau\frac{\sigma_{s}^{2}}{4}\partial_{x}^{2}J.
\end{multline}
Being $\operatorname{Re}\langle\Psi|U_{\tau}^{\dag}|x_{s}\rangle\langle
x_{s}|U_{\tau}X|\Psi\rangle=x_{s}\rho-\tau J$, and substituting Eq. (\ref{1}) and
(\ref{2}) into Eq. (\ref{epss}), we finally have
\begin{multline}
\varepsilon_{s}(x_{s})=\\=\tau^{-1}\left\vert \dfrac{4x_{s}\rho+2\sigma_{s}%
^{2}\partial_{x}\rho+x_{s}\sigma_{s}\partial_{x}^{2}\rho-4\tau J-\tau
\sigma_{s}^{2}\partial_{x}^{2}J}{4\rho+\sigma_{s}^{2}\partial_{x}^{2}\rho
}-\right.\\ \left.-\dfrac{x_{s}\rho-\tau J}{\rho}\right\vert =\\
=\tau^{-1}\left\vert \dfrac{2\sigma_{s}^{2}\partial_{x}\rho+\tau v\sigma
_{s}^{2}\partial_{x}^{2}\rho-\tau\sigma_{s}^{2}\partial_{x}^{2}J}{4\rho
+\sigma_{s}^{2}\partial_{x}^{2}\rho}\right\vert =\\=\sigma_{s}^{2}\left\vert
\dfrac{\frac{2}{\tau}\partial_{x}\rho+\left(  2\partial_{x}\rho-\rho
\partial_{x}\right)  \partial_{x}v}{4\rho+\sigma_{s}^{2}\partial_{x}^{2}\rho
}\right\vert
\end{multline}

\section{Derivation of the error $\varepsilon_{w}(x_{s})$}
\label{appendixC}

The definition of $\varepsilon_{w}(x_{s})$ is:
\begin{multline}
\varepsilon_{w}(x_{s})=\tau^{-1}\left\vert \dfrac{\int dx_{w}x_{w}\langle
\Psi|\hat{W}_{w}^{\dag}U_{\tau}^{\dag}\hat{S}_{s}^{\dag}\hat{S}_{s}U_{\tau
}\hat{W}_{w}\Psi\rangle}{\int dx_{w}\langle\Psi|\hat{W}_{w}U_{\tau}^{\dag}%
\hat{S}_{s}^{\dag}\hat{S}_{s}U_{\tau}\hat{W}_{w}|\Psi\rangle}-\right.\\-\left.\dfrac
{\operatorname{Re}\langle\Psi|U_{\tau}^{\dag}\hat{S}_{s}^{\dag}\hat{S}%
_{s}U_{\tau}\hat{x}|\Psi\rangle}{\langle\Psi|U_{\tau}^{\dag}\hat{S}_{s}^{\dag
}\hat{S}_{s}U_{\tau}|\Psi\rangle}\right\vert .\label{espw}%
\end{multline}
Under the limit (11) and after the integration over $x_{w}$ we can expand $\exp\left[  -\left(
x^{\prime\prime}-x^{\prime}\right)  ^{2}/4\sigma_{w}^{2}\right]$ in Taylor series
in the numerator and denominator of (\ref{espw}) to get
\begin{multline}
\label{ap01}
\int dx_{w}\langle\Psi|\hat{W}_{w}^{\dag}U_{\tau}^{\dag}\hat{S}_{s}^{\dag}%
\hat{S}_{s}U_{\tau}\hat{W}_{w}\Psi\rangle=
\langle\Psi|U_{\tau}^{\dag}\hat{S}_{s}^{\dag}\hat{S}_{s}U_{\tau}|\Psi
\rangle-\\-\frac{1}{2\sigma_{w}^{2}}\left(  \operatorname{Re}\langle\Psi|U_{\tau
}^{\dag}\hat{S}_{s}^{\dag}\hat{S}_{s}U_{\tau}\hat{x}^{2}|\Psi\rangle
-\langle\Psi|\hat{x}U_{\tau}^{\dag}\hat{S}_{s}^{\dag}\hat{S}_{s}U_{\tau}%
\hat{x}|\Psi\rangle\right)
\end{multline}
and
\begin{multline}
\int dx_{w}x_{w}\langle\Psi|\hat{W}_{w}^{\dag}U_{\tau}^{\dag}\hat{S}_{s}%
^{\dag}\hat{S}_{s}U_{\tau}\hat{W}_{w}\Psi\rangle=\label{ap02}\\
=\operatorname{Re}\langle\Psi|U_{\tau}^{\dag}\hat{S}_{s}^{\dag}\hat{S}%
_{s}U_{\tau}\hat{x}|\Psi\rangle-\frac{1}{4\sigma_{w}^{2}}\times\\ \times\left(
\operatorname{Re}\langle\Psi|U_{\tau}^{\dag}\hat{S}_{s}^{\dag}\hat{S}%
_{s}U_{\tau}\hat{x}^{3}|\Psi\rangle-\operatorname{Re}\langle\Psi|\hat
{x}U_{\tau}^{\dag}\hat{S}_{s}^{\dag}\hat{S}_{s}U_{\tau}\hat{x}^{2}|\Psi
\rangle\right)  .
\end{multline}
Moreover using twice Eq. \pref{commutator} we have
\begin{multline}
\langle\Psi|\hat{x}U_{\tau}^{\dag}\hat{S}_{s}^{\dag}\hat{S}_{s}U_{\tau}\hat
{x}|\Psi\rangle=\operatorname{Re}\left(  \langle\Psi|U_{\tau}^{\dag}\hat
{S}_{s}^{\dag}\hat{S}_{s}U_{\tau}\hat{x}^{2}|\Psi\rangle+\right. \\ \left. +\frac{\tau}{m}%
\langle\Psi|U_{\tau}^{\dag}[\hat{S}_{s}^{\dag}\hat{S}_{s},\hat{p}]U_{\tau}%
\hat{x}|\Psi\rangle\right)  \label{ap1}\\
\end{multline}
and
\begin{multline}
\operatorname{Re}\langle\Psi|\hat{x}U_{\tau}^{\dag}\hat{S}_{s}^{\dag}\hat
{S}_{s}U_{\tau}\hat{x}^{2}|\Psi\rangle=\operatorname{Re}\left(  \langle
\Psi|U_{\tau}^{\dag}\hat{S}_{s}^{\dag}\hat{S}_{s}U_{\tau}\hat{x}^{3}%
|\Psi\rangle+\right. \\ \left. +\frac{\tau}{m}\langle\Psi|U_{\tau}^{\dag}[\hat{S}_{s}^{\dag}%
\hat{S}_{s},\hat{p}]U_{\tau}\hat{x}^{2}|\Psi\rangle\right)  .\label{ap2}%
\end{multline}
Putting Eq. (\ref{commutator_appendix}) into Eqs. (\ref{ap1}) and (\ref{ap2}) and
substituting them into Eqs. (\ref{ap01}) and (\ref{ap02}) we have
\begin{multline}
\int dx_{w}\langle\Psi|\hat{W}_{w}^{\dag}U_{\tau}^{\dag}\hat{S}_{s}^{\dag}%
\hat{S}_{s}U_{\tau}\hat{W}_{w}\Psi\rangle=\\=\langle\Psi|U_{\tau}^{\dag}\hat
{S}_{s}^{\dag}\hat{S}_{s}U_{\tau}|\Psi\rangle+\frac{\tau\hbar}{2m\sigma
_{w}^{2}}\partial_{x_{s}}\operatorname{Im}\langle\Psi|U_{\tau}^{\dag}\hat
{S}_{s}^{\dag}\hat{S}_{s}U_{\tau}\hat{x}|\Psi\rangle
\end{multline}
and%
\begin{multline}
\int dx_{w}x_{w}\langle\Psi|\hat{W}_{w}^{\dag}U_{\tau}^{\dag}\hat{S}_{s}%
^{\dag}\hat{S}_{s}U_{\tau}\hat{W}_{w}\Psi\rangle=\\=\operatorname{Re}\langle
\Psi|U_{\tau}^{\dag}\hat{S}_{s}^{\dag}\hat{S}_{s}U_{\tau}X|\Psi\rangle
+\\+\frac{\tau\hbar}{4m\sigma_{w}^{2}}\partial_{x_{s}}\operatorname{Im}%
\langle\Psi|U_{\tau}^{\dag}\hat{S}_{s}^{\dag}\hat{S}_{s}U_{\tau}\hat{x}%
^{2}|\Psi\rangle.
\end{multline}
Using again Eqs. \pref{commutator} and (\ref{commutator_appendix}) we realize that%
\begin{align}
\operatorname{Im}\langle\Psi|U_{\tau}^{\dag}\hat{S}_{s}^{\dag}\hat{S}%
_{s}U_{\tau}\hat{x}|\Psi\rangle & =\frac{\hbar\tau}{2m}\partial_{x_{s}}%
\langle\Psi|U_{\tau}^{\dag}\hat{S}_{s}^{\dag}\hat{S}_{s}U_{\tau}|\Psi\rangle\\
\operatorname{Im}\langle\Psi|U_{\tau}^{\dag}\hat{S}_{s}^{\dag}\hat{S}%
_{s}U_{\tau}\hat{x}^{2}|\Psi\rangle & =\frac{\hbar\tau}{m}\partial_{x_{s}%
}\operatorname{Re}\langle\Psi|U_{\tau}^{\dag}\hat{S}_{s}^{\dag}\hat{S}%
_{s}U_{\tau}\hat{x}|\Psi\rangle
\end{align}
so finally we can write%
\begin{multline}
\int dx_{w}\langle\Psi|\hat{W}_{w}^{\dag}U_{\tau}^{\dag}\hat{S}_{s}^{\dag}%
\hat{S}_{s}U_{\tau}\hat{W}_{w}\Psi\rangle  =\\=\left(  1+\frac{\tau^{2}\hbar
^{2}}{4m^{2}\sigma_{w}^{2}}\partial_{x_{s}}^{2}\right)  \langle\Psi|U_{\tau
}^{\dag}\hat{S}_{s}^{\dag}\hat{S}_{s}U_{\tau}|\Psi\rangle\label{ap21}
\end{multline}
and
\begin{multline}
\int dx_{w}x_{w}\langle\Psi|\hat{W}_{w}^{\dag}U_{\tau}^{\dag}\hat{S}_{s}%
^{\dag}\hat{S}_{s}U_{\tau}\hat{W}_{w}\Psi\rangle  =\\=\left(  1+\frac{\tau
^{2}\hbar^{2}}{4m^{2}\sigma_{w}^{2}}\partial_{x_{s}}^{2}\right)
\operatorname{Re}\langle\Psi|U_{\tau}^{\dag}\hat{S}_{s}^{\dag}\hat{S}%
_{s}U_{\tau}\hat{x}|\Psi\rangle\label{ap22}%
\end{multline}
Evaluating the derivatives in (\ref{ap21}) and (\ref{ap22}), we have
\begin{multline}
\partial_{x_{s}}^{2}\langle\Psi|U_{\tau}^{\dag}\hat{S}_{s}^{\dag}\hat{S}%
_{s}U_{\tau}|\Psi\rangle=C_{s}^{2}\partial_{x_{s}}^{2}\int e^{-\frac{\left(
x_{s}-x\right)  ^{2}}{\sigma_{s}^{2}}}\rho(x)dx=\\=-C_{s}^{2}\frac{4}{\sigma
_{s}^{4}}\int e^{-\frac{\left(  x_{s}-x\right)  ^{2}}{\sigma_{s}^{2}}}\left(
-(x_{s}-x)^{2}+\frac{\sigma_{s}^{2}}{2}\right)  \rho(x)dx\label{ap31}%
\end{multline}
and
\begin{multline}
\partial_{x_{s}}^{2}\operatorname{Re}\langle\Psi|U_{\tau}^{\dag}S^{\dag
}SU_{\tau}X|\Psi\rangle=\\=-C_{s}^{2}\frac{4}{\sigma_{s}^{4}}\int e^{-\frac
{\left(  x_{s}-x\right)  ^{2}}{\sigma_{s}^{2}}}\left(  -(x_{s}-x)^{2}%
+\frac{\sigma_{s}^{2}}{2}\right) \times\\ \times\left(  x\rho(x)-\tau J(x)\right)
dx\label{ap32}%
\end{multline}
which, both can be rewritten in a compact way as
\begin{multline}
-C_{s}^{2}\frac{4}{\sigma_{s}^{2}}\int e^{-\frac{\left(  x_{s}-x\right)  ^{2}%
}{\sigma_{s}^{2}}}\left(  -(x_{s}-x)^{2}+\frac{\sigma_{s}^{4}}{2}\right)
\alpha(x)dx \approx \\ \approx \partial_{x_{s}}^{2}\alpha(x_{s})\label{der}%
\end{multline}
where we keep only the first three terms in the Taylor expansion.
Using Eq. (\ref{der}) in Eqs. (\ref{ap21}) and (\ref{ap22}) and plugging them into expression (\ref{espw}) we have%
\begin{multline}
\varepsilon(\sigma_{w})=\\=\tau^{-1}\frac{\tau^{2}\hbar^{2}}{4m^{2}\sigma_{w}%
^{2}}\left\vert \dfrac{\partial_{x}^{2}\left(  x\rho-\tau J\right)
-\tfrac{\operatorname{Re}\langle\Psi|U_{\tau}^{\dag}\hat{S}_{s}^{\dag}\hat
{S}_{s}U_{\tau}\hat{x}|\Psi\rangle}{\langle\Psi|U_{\tau}^{\dag}\hat{S}%
_{s}^{\dag}\hat{S}_{s}U_{\tau}|\Psi\rangle}\partial_{x}^{2}\rho}{\langle
\Psi|U_{\tau}^{\dag}\hat{S}_{s}^{\dag}\hat{S}_{s}U_{\tau}|\Psi\rangle
+\frac{\tau^{2}\hbar^{2}}{4m^{2}\sigma_{w}^{2}}\partial_{x}^{2}\rho
}\right\vert
\end{multline}
which can be finally rewritten using equations (\ref{1}) and (\ref{2}) as
\begin{multline}
\varepsilon(\sigma_{w})=\\=\frac{\tau\hbar^{2}}{4m^{2}\sigma_{w}^{2}}\left\vert
\dfrac{2\partial_{x}\rho-\tau\partial_{x}^{2}J-\tfrac{2\sigma_{s}^{2}%
\partial_{x}^{2}\rho-4\tau J-\tau\sigma_{s}^{2}\partial_{x}^{2}J}{4\rho
+\sigma_{s}^{2}\partial_{x}^{2}\rho}\partial_{x}^{2}\rho}{\rho+\frac
{\sigma_{s}^{2}}{4}\partial_{x}^{2}\rho+\frac{\tau^{2}\hbar^{2}}{4m^{2}%
\sigma_{w}^{2}}\partial_{x}^{2}\rho}\right\vert .
\end{multline}
In the limit of small $\sigma_{s}$ we finally get%
\begin{multline}
\varepsilon(\sigma_{w})=\\=\frac{\tau\hbar^{2}}{4m^{2}\sigma_{w}^{2}}\left\vert
\dfrac{2\partial_{x}\rho-\tau\partial_{x}^{2}J+\tau v\partial_{x}^{2}\rho
}{\rho+\frac{\tau^{2}\hbar^{2}}{4m^{2}\sigma_{w}^{2}}\partial_{x}^{2}\rho
}\right\vert =\\=\frac{\tau\hbar^{2}}{4m^{2}\sigma_{w}^{2}}\left\vert
\dfrac{2\left(  1-\tau\partial_{x}v\right)  \partial_{x}\rho-\tau\rho
\partial_{x}^{2}v}{\rho+\frac{\tau^{2}\hbar^{2}}{4m^{2}\sigma_{w}^{2}}%
\partial_{x}^{2}\rho}\right\vert .
\end{multline}

\end{document}